\documentclass{article}

\newcommand{\bm}[1]{\boldsymbol{#1}}
\newcommand{\ba}[1]{\mathbf{#1}}              
\newcommand{\beq}[0]{\begin{eqnarray}}
\newcommand{\eeq}[0]{\end{eqnarray}}
\newcommand{\beqn}[0]{\begin{eqnarray*}}
\newcommand{\eeqn}[0]{\end{eqnarray*}}
\newcommand{\pp}[2]{\frac{\partial #1}{\partial #2}}  

\usepackage{tabulary}
\usepackage{color}
\usepackage[table]{xcolor}
\usepackage{multirow}
\usepackage{amsfonts,amsmath,amssymb}
\usepackage{graphicx}
\usepackage{caption}
\usepackage{tabulary}
\usepackage[]{siunitx}
\usepackage{authblk}
\usepackage{natbib}

\begin{document}
	
\title{Non-invasive in silico determination of ventricular wall pre-straining and characteristic cavity pressures}
	
\author[1]{Sebastian Skatulla}
\author[2]{Carlo Sansour}
\author[1,3]{Mary Familusi}
\author[4]{Jagir Hussan}
\author[5,6,7,8]{Ntobeko Ntusi}

\affil[1]{Computational Continuum Mechanics Research Group, Department of Civil Engineering, University of Cape Town, South Africa}
\affil[2]{Department of Mathematics, Bethlehem University, Bethlehem,  Palestine}
\affil[3]{South African DST-NRF Centre of Excellence in Epidemiological Modelling and Analysis, Stellenbosch University, South Africa}
\affil[4]{Auckland Bioengineering Institute, University of Auckland, New Zealand}
\affil[5]{Division of Cardiology, Department of Medicine, University of Cape Town, Cape Town, South Africa}
\affil[6]{Cape Heart Institute, Faculty of Health Sciences, University of Cape Town, South Africa}
\affil[7]{Cape Universities Body Imaging Centre, Faculty of Health Sciences, University of Cape Town, South Africa}
\affil[8]{South African Medical Research Council Extramural Unit on Intersection of Noncommunicable Diseases and Infectious Diseases}

\maketitle	
	
\begin{abstract}
The clinical application of patient-specific modelling of the heart can provide valuable insights in supplementing and advancing methods of diagnosis as well as helping to devise the best possible therapeutic approach for each individual pathological heart condition. The potential of computational cardiac mechanics, however, has not yet been fully leveraged due to the heart's complex physiology and limitations in the non-invasive in vivo characterisation of heart properties necessary required for accurate patient-specific modelling such as the heart anatomy in an unloaded state, ventricular pressure, the elastic constitutive parameters and the myocardial muscle fibre orientation distribution. From a solid mechanics point of view without prior knowledge of the unloaded heart configuration and the cavity pressure-volume evolution, in particular, the constitutive parameters cannot be accurately estimated to describe the highly nonlinear elastic material behaviour of myocardial tissue.
Here, knowledge of the volume-normalized end-diastolic pressure relation for larger mammals is exploited in combination with a novel iterative inverse parameter optimisation framework to determine end-systolic and end diastolic pressures, ventricular wall pre-straining and pre-stressing due the residual end-systolic cavity pressure as well as myocardial tissue stiffness parameters for biventricular heart models. 
\end{abstract}
	



%
\section{Introduction}
Realistic modelling of the whole beating heart is the ”holy grail” of computational cardiac mechanics which has been achieved only recently by the international “Living Heart Project” research initiative \cite{baillargeon2014living}. The objective of patient-specific modelling of the heart, however, is still out of reach, because further intense research efforts are necessary to determine the biomechanical properties of healthy and diseased heart muscle tissue in-vivo, meaning non-invasive. The effective implementation and evaluation of the clinical use of patient-specific modelling supplementing personalized therapy is mostly unknown and under debate \cite{gray2018patient}. In patient-specific heart modelling, it is crucial to be able to calibrate the various model parameters to clinically observed data, because models can only be relied on and be used for prediction if an observed physiological state can be represented.

Patient-specific biomechanical modelling of the heart has to account for the presence of a physiological pressure load as a result of the  pre-stressed state of the imaged tissue. Two main sources of pre-stressing need to be considered: Firstly, cardiac tissue exhibits growth- and remodelling-related residual stresses even if completely unloaded ex vivo \cite{omens2003complex,shi2019epicardial,zhuan2022volumetric} and secondly, anatomical heart models are usually based on in vivo imaging of the beating heart when exhibiting the lowest cavity pressures at end systole, see e.g. \cite{nikou2016computational}. The latter implies that the heart tissue is still experiencing stress, strain and deformation due to the remaining pressure. Failure to account for this pre-stressed state in solid tissue mechanics models results in inaccurate metrics, which are then used for health evaluation, risk assessment, or surgical planning. 

In solid mechanics, the load application and corresponding deformation, strain and stress response is analyzed as linked to the unloaded material configuration of the problem at hand. This requires knowledge of the unloaded configuration which is in cardiac mechanics typically unknown. However, with the knowledge of the residual cavity pressure, the unknown unloaded configuration of the heart can be found by solving an inverse problem based on its known pre-stressed configuration. This inverse problem can usually only be solved computationally, in particular for biological tissue mechanics applications due to the highly nonlinear nature. \citet{rajagopal2007determining} proposed a incremental and iterative method applying a finite difference approximation of the finite element (FE) residual vector formulated in the Lagrangian form to predicted the undeformed shape of human breasts. \citet{bols2013computational} proposed backward displacement method making use of a fixed point algorithm to iteratively retrieve the in vivo stress in a mouse-specific abdominal aorta due to the blood pressure at the moment of imaging and to reinstate the zero-pressure blood vessel geometry. Using in-vivo pressure data and magnetic resonance imaging, \citet{pourmodheji2021inverse} applied a patient-specific inverse FE modelling framework to iterative determine besides of the elastic material parameters the pre-stretch of elastin and collagen fibres in the proximal pulmonary arteries in order to assess mechanical and structural alterations of these micro-constituents. 
In contrast, a direct approach requires the governing equations to be consistently related to the known deformed configuration. Accordingly, \citet{govindjee1996computational} and \citet{gee2010computational} derived the corresponding weak form as integrals over the current body. \citeauthor{govindjee1996computational} also elaborated on the difficulty concerning a conservative traction boundary condition when related to quantities defined in current configuration. Applying this approach to a biventricular heart model, it was shown that the inclusion of pre-stressing leads to significant strain contributions in the myocardium \cite{peirlinck2018modular}. In the work of \citet{hirschvogel2017monolithic} the imaging-based patient-specific ventricular model was pre-stressed to low end-diastolic pressure to account for the imaged, stressed configuration, using a modified updated Lagrangian FE formulation proposed by \citet{gee2010computational} where on element level only the deformation gradient and the isoparametric Jacobian matrix was updated.

The modelling of the deformation response of biological tissue to the applied loading additionally requires the determination of the parameters of the nonlinear elastic material relations.
Initially, patient-specific mechanical properties of blood arteries were estimated by analytical equations such as the Young-Laplace Law. However, the underlying assumptions of material homogeneity and isotropy contradicts the known complexity of myocardial and vascular tissue \cite{moulton1995inverse}. \citet{klotz2006single} uncovered from ex vivo and in vivo measurements of end diastolic pressure volume relations (EDPVR) and subsequent regression analysis that normalized EDPVR curves of the left ventricle (LV) had a consistent profile across humans and a variety of mammals, regardless of etiology. Using the least-squared error between the so-called "Klotz curve" and the modelled ventricular pressure-volume curve as an optimization objective increases the number of comparison points and facilitates the fitting of non-linear myocardial material models, e.g. \cite{genet2014distribution,palit2018vivo,sack2018construction}. The use of the "Klotz curve" requires at least on measured pressure-volume pair which is usually the end diastolic pressure and volume. However, in cardiac mechanics specifically, the in vivo patient-specific LV pressure is often not known as well, because the only accurate method for LV pressure measurement is invasive catheterization which is therefore not readily available on an individual basis. Without prior knowledge of the so-called end-diastolic pressure-volume relation, it is effectively not possible to calibrate the anisotropic highly non-linear elastic material relations for myocardial tissue. 

To address the problem when direct pressure measurements are unavailable, this work proposes to combine a direct inverse method to determine the unloaded configurations of biventricular patient-specific heart models with the previously mentioned "Klotz curve" and the bounded Levenberg-Marquart parameter optimisation method \cite{}. This allows for the simultaneous computation of the unknown end-diastolic ventricular pressures, pre-stressing and elastic material constants.
In contrast to the related inverse unloaded configuration determination approaches by \citet{govindjee1996computational} and \citet{gee2010computational}, the variational principle in the known deformed configuration is derived from the conventional Lagrangian formulation.

The paper is organized as follows:  In Sec.~\ref{standardTheorySection} the basics of a standard total Lagrangian approach are revised. Based on this, a non-standard total Lagrangian approach is introduced in Sec.~\ref{nonstandardTheorySection} which allows for direct computation of the unloaded configuration based on a know loaded configuration. Finally, in Sec.~\ref{examplesSection} the proposed method is applied to three numerical examples modelling non-linear hyperelastic material.

\section{Standard total Lagrangian approach}
\label{standardTheorySection}

This section provides only in brief the basic principles of a total Lagrangian approach
including necessary kinematical assumptions, variational principle and 
its governing equations. For further details the reader is referred to e.g. \cite{Gurtin2010}.

\subsection{Kinematics}\label{kinematicsSection}

Let $\mathbb{E}(3)$ be the Euclidian vector space and $\mathcal{B}\subset\mathbb{E}(3)$,
where $\mathcal{B}$ is a three-dimensional manifold defining a material body. A motion of
$\mathcal{B}$ is represented by a one parameter non-linear deformation mapping $\bm\varphi_t:\mathcal{B}
\rightarrow\mathcal{B}_t$, where $t\in\mathbb{R}$ is the time and $\mathcal{B}_t$ is the current
configuration at time $t$. Accordingly, each material point $\ba X\in\mathcal{B}$ is related
to its placement $\ba x$ in the spatial configuration
$\mathcal{B}_t$ by the mapping
\begin{eqnarray}
\ba x = \bm\varphi\left(\ba X,t\right)
\label{macroMap}
\end{eqnarray}
In what follows and without loss of generality  we identify the body $\mathcal{B}$ with its 
undeformed reference configuration at a fixed time $t_0$. The deformation map possesses an 
invertible linear tangent map $\ba F = \mbox{Grad}\,\bm\varphi$ denoted by the deformation 
gradient, where the Jacobian $J = \mbox{det}\,\ba F > 0$. The operator
$\mbox{Grad}$ represents the gradient with respect to the reference configuration
\begin{eqnarray}
\mbox{Grad} := \frac{\partial}{\partial\ba X}\,.
\end{eqnarray}
The body $\mathcal{B}$ is parameterized by the Cartesian coordinates $X_i,\,i = 1,\,2,\,3$.
Here, and in what follows, Latin indices take the values $1,\,2\,\,\mbox{or}\,\,3$.
As a deformation measure we make use of the right {\it Cauchy-Green} deformation tensor $\bf C$ defined by
\beq
\ba C = \ba F^T \ba F\,.\label{rightCauchyGreen}
\eeq
In the following we want to confine ourselves to the quasi static case.

\subsection{Variational principle}

Let us consider a non-linear boundary value problem on domain $\mathcal{B}$
with boundary $\partial\mathcal{B}$. Dirichlet boundary conditions are prescribed
on $\partial\mathcal{B}_D\subset\partial\mathcal{B}$ and Neumann boundary conditions
are prescribed on $\partial\mathcal{B}_N = \partial\mathcal{B}\setminus\partial\mathcal{B}_D$.
We define the external virtual work in the Lagrangian form as follows
\begin{eqnarray}
\mathcal{W}_{ext} = \int_\mathcal{B}\ba b\cdot\delta\ba u\,\,dV 
+\int_{\partial\mathcal{B}_N} 
\ba t^{(\ba n)}\cdot\delta\ba u\,\,dA
\label{standardExternalWork}
\end{eqnarray}
where $\ba b$ denotes the  body force, $\ba t^{(\ba n)}$ the external traction and $\ba n$ 
the outward normal on $\partial\mathcal{B}_N$.

Furthermore, we assume now that the body under consideration 
$\mathcal{B}$ is hyperelastic and possesses an elastic potential $\Psi$ 
represented by the stored strain energy per unit mass $\psi(\ba C)$.
The variation of the internal potential with respect to
$\ba C$ in the Lagrangian form reads as follows
\begin{eqnarray}
\delta\Psi = \int_{\mathcal{B}}
\,\rho_0\,\frac{\partial\psi} {\partial\ba C}:\delta\ba C\,\,dV\,.
\label{standardInternalPower}
\end{eqnarray}
In the static case and by considering only mechanical processes
the {\it first law of thermodynamics} provides the following
variational statement
\begin{eqnarray}
\delta\Psi - \mathcal{W}_{ext} = 0\,.
\end{eqnarray}
Upon substitution of Eqs.~(\ref{standardExternalWork}) and (\ref{standardInternalPower})
this variational statement can be written as
\begin{eqnarray}
\int_{\mathcal{B}}
\frac{1}{2}\,\ba S:\delta\ba C\,\,dV -\int_\mathcal{B}\ba b\cdot\delta\ba u\,\,dV -
\int_{\partial\mathcal{B}_N}\,\ba t\cdot\delta\ba u\,\,dA = 0\,,
\label{standardVariationalPrinciple}
\end{eqnarray}
with the second {\it Piola-Kirchhoff} stress tensor $\ba S$ given by
\beq
\ba S = 2\rho_0\frac{\partial\psi}{\partial\ba C}\,.
\eeq
The governing equations of the above variational formulation (Eq.~\ref{standardVariationalPrinciple}) 
can be derived by applying {\it Gauss's divergence theorem} and, assuming the variation
$\delta\ba u = \ba 0$ on $\partial\mathcal{B}_D$ but otherwise arbitrary, we have
\beq
\mbox{Div}\left(\ba F\ba S\right) + \ba b = \ba 0\quad\mbox{on}\,\,\mathcal{B}\,,
\eeq
and the corresponding Neumann boundary condition is given by
\beq
\ba F\ba S\ba n = \ba t^{\left(\ba n\right)}
\quad\mbox{on}\,\,\partial\mathcal{B}_N\,.
\eeq
where Div denotes the divergence operator in the reference configuration.
These field equations are supplemented by essential boundary conditions, the so-called
Dirichlet boundary conditions
\begin{eqnarray}
\ba u = \ba g\quad\mbox{on}\,\,\partial\mathcal{B}_D\,.
\end{eqnarray}
with $\ba g$ being the displacement field prescribed on $\partial\mathcal{B}_D$.

%
\section{Non-standard total Lagrangian approach}
\label{nonstandardTheorySection}

If the unloaded and stress-free configuration $\mathcal{B}$ is not known but only 
a certain deformed configuration $\hat{\mathcal{B}}$, the standard total Lagrangian
approach as introduced in the previous section cannot be directly used but needs
to be modified. The reason for this is found in the fact that in the total 
Lagrangian approach the displacement field and its gradient as well as stress 
and strain quantities refer to the reference configuration which is unknown.

\begin{figure}[t!]
\centering
\includegraphics*[width=70mm]{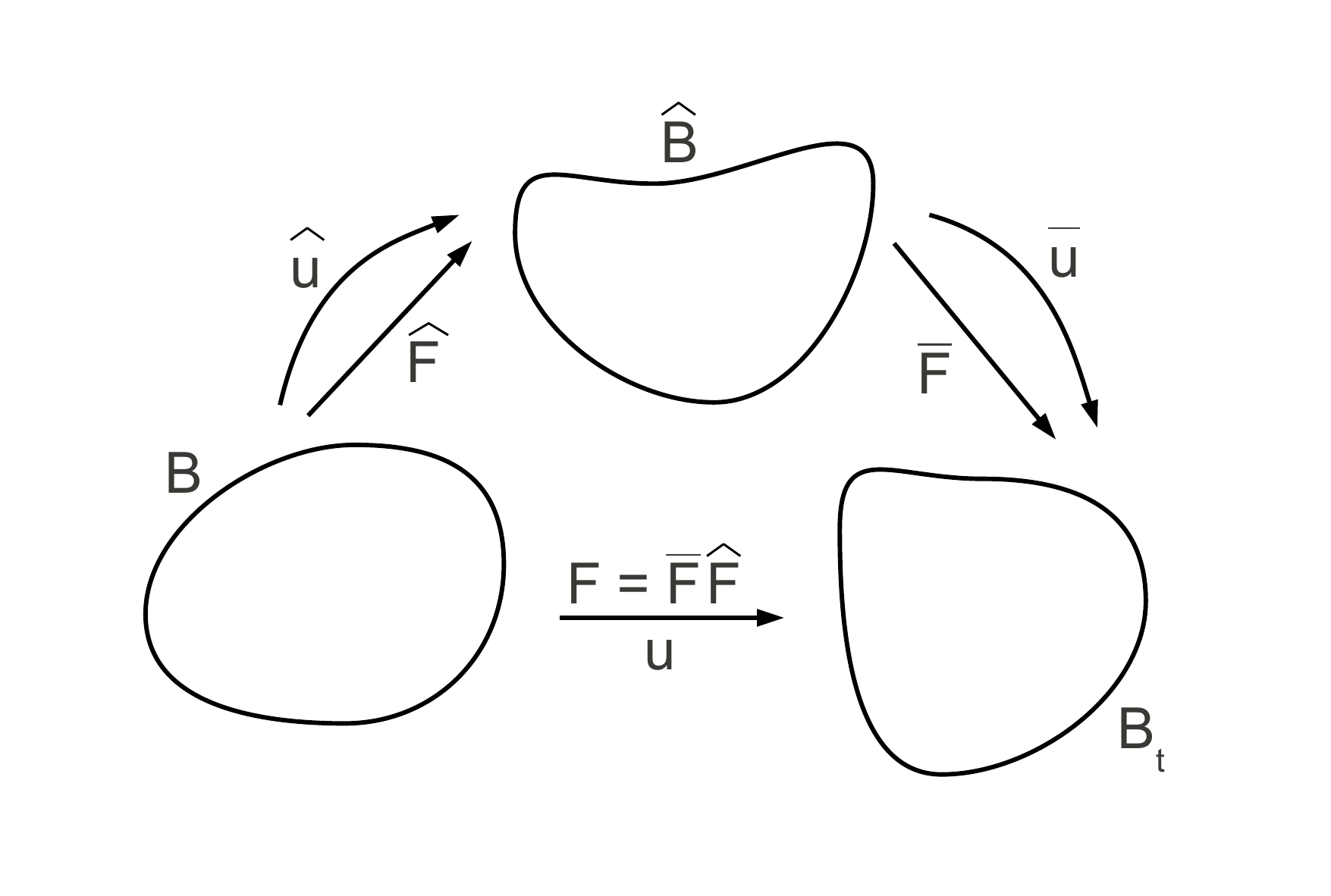}
\caption{Configuration spaces.}
\label{configuration-spaces}
\end{figure}
In order to proceed let us first introduce some deformation maps: As proposed by \citet{gee2010computational} the unknown reference 
configuration $\mathcal{B}$ is linked with the know deformed configuration
$\hat{\mathcal{B}}$ by $\hat{\ba F}$ and any subsequent deformed configuration 
$\mathcal{B}_t$ by $\bar{\ba F}$. Thus, 
\beq
\ba F = \bar{\ba F}\,\hat{\ba F} = \pp{\ba x}{\hat{x}_i}\pp{\hat{x}_i}{\ba X}
\label{totalDefgrad}
\eeq
relates $\mathcal{B}$ with $\mathcal{B}_t$ as illustrated in Fig.~\ref{configuration-spaces}. 

\subsection{Inverse problem} 
\label{inverseProblemSection}
First, the objective is to find 
\beq
\hat{\ba F} = \pp{\hat{\ba x}}{\ba X}
\label{hatDefgrad}
\eeq
with $\hat{\ba x}\in\hat{\mathcal{B}}$ being known which we want to call
{\it "inverse problem"}. For this 
the standard variational principle needs to be reformulated such as 
to determine $\ba X\in\mathcal{B}$ by employing the inverse of the tangent map 
(Eq.~\ref{hatDefgrad})
\beq
\hat{\ba F}^{-1} = \pp{(\hat{\ba x}-\hat{\ba u})}{\hat{\ba x}}\,,
\label{hatDefgradInv}
\eeq
where the displacement field $\hat{\ba u}$ translates $\mathcal{B}$ to the 
$\hat{\mathcal{B}}$.

Let us start with the standard variational formulation in its Lagrangian form 
(Eq.~\ref{standardVariationalPrinciple}) evaluated when $\mathcal{B}$ is deformed
into $\hat{\mathcal{B}}$ such that it holds $\ba F = \hat{\ba F}$:
\beq
\int_{\mathcal{B}}\,\hat{\ba F}\ba S:\delta\hat{\ba F}\,dV 
-\mathcal{W}_{ext} = 0
\label{inverseVariationalPrinciple1}
\eeq
where the second {\it Piola-Kirchhoff} stress tensor is now a function 
of $\hat{\ba C} = \hat{\ba F}^T\hat{\ba F}$, i.e.
\beq
\ba S = 2\rho_0\frac{\partial\psi(\hat{\ba C})}{\partial\hat{\ba C}}\,.
\eeq
From
\beqn
\hat{\ba F}\hat{\ba F}^{-1} = \ba 1
\eeqn
which is varied with respect to $\hat{\ba u}$
\beqn
\delta\hat{\ba F}\hat{\ba F}^{-1} + \hat{\ba F}\delta\hat{\ba F}^{-1} = \ba 0
\eeqn
we obtain
\beq
\delta\hat{\ba F} = -\hat{\ba F}\delta\hat{\ba F}^{-1}\hat{\ba F}\,.
\eeq
This is substituted together with $dV = \hat{J}^{-1}d\hat{v}$ into the above standard variational 
formulation (Eq.~\ref{inverseVariationalPrinciple1}) and we have
\beq
\int_{\hat{\mathcal{B}}}\,-\hat{J}^{-1}\hat{\ba C}\hat{\ba S}\hat{\ba F}^T:
\delta\hat{\ba F}^{-1}\,d\hat{v} -\mathcal{W}_{ext} = 0\,,
\label{inverseVariationalPrinciple2}
\eeq
which is with Eq.~\eqref{hatDefgradInv} and the Cauchy stress tensor given by $\hat{\bm\sigma} = -\hat{J}^{-1}\hat{\ba C}\hat{\ba S}\hat{\ba F}^T$ analog to \cite{gee2010computational,govindjee1996computational,peirlinck2018modular}.
Here, it is important to note that, as the integration needs to be performed 
over $\hat{\mathcal{B}}$, the approximation of geometry refers to 
$\hat{\ba x}\in\hat{\mathcal{B}}$ and corresponding spatial derivatives 
$\displaystyle\pp{}{\hat{\ba x}}\left(\bullet\right)$. 

In terms of loading, we need discriminate between conservative and non-conservative loading, i.e.
whether it depends on deformation using the standard total Lagriangian approach
(Eq.~\ref{standardVariationalPrinciple}). For non-conservative loading, e.g. pressure loading $p$, 
using {\it Nanson's} formula $\hat{\ba n}\,d\hat{a} = \hat{J}\hat{\ba F}^{-T}\ba n\,dA$
which establishes a relation between known deformed and corresponding unknown undeformed 
surface elements $d\hat{\ba a}$ and $d\ba A$, respectively,
the external virtual work is expressed in the unknown undeformed configuration
as
\beq
\mathcal{W}_{ext} = 
\int_{\partial\mathcal{B}_N}\,p\,\hat{J}\hat{\ba F}^{-T}\ba n\cdot\delta\hat{\ba u}\,dA
\eeq
whereas in the known deformed configuration as 
\beq
\mathcal{W}_{ext} = 
\int_{\partial\hat{\mathcal{B}}_N}\,p\,\hat{\ba n}\cdot\delta\hat{\ba u}\,d\hat{a}
\label{intermediateExternalPressure}
\eeq
with $\hat{\ba n}$ denoting the normal vector on $\partial\hat{\mathcal{B}}$
which does not depend on the deformation. The latter is a crucial point 
to realize, as it is in contrast to standard total Lagriangian approach. 
When dealing with the inverse problem 
$\hat{\mathcal{B}}$ is assumed fixed whereas $\mathcal{B}_0$ needs to be 
incrementally approached due to the nonlinear nature of the problem. 
Consequently, when solving the the variational problem 
(Eq.~\ref{inverseVariationalPrinciple2}) numerically, e.g. employing the 
Newton-Raphson method, the pressure does not appear in the variational statement's linearized form. 
Conversely, when applying a conservative loading, e.g. a traction force acting 
on the undeformed surface $\partial\mathcal{B}_N$, we have
\beq
\mathcal{W}_{ext} = 
\int_{\partial\mathcal{B}_N}\,\ba t\cdot\delta\hat{\ba u}\,dA
\eeq
or with respect to the known deformed surface $\partial\hat{\mathcal{B}}_N$ 
\beq
\int_{\partial\hat{\mathcal{B}}_N}
\,\ba t\cdot\delta\hat{\ba u}\,
\hat{J}^{-1}\,|\hat{\ba F}^{T}\,\hat{\ba n}|\,d\hat{a}
\label{inverseExternalTraction}
\eeq
where the sizes of undeformed and deformed surface elements can be obtained
form {\it Nanson's} formula 
$|\ba n|\,dA = \hat{J}^{-1}\,|\hat{\ba F}^{T}\,\hat{\ba n}|\,d\hat{a}$.
Consequently, for inverse problem the linearization of the surface traction
contribution (Eq.~\ref{inverseExternalTraction}) does not vanish.


\subsection{Forward problem}
\label{forwardProblemSection}

Once the map $\hat{\ba F}$ linking unknown undeformed configuration $\mathcal{B}_0$
and known deformed configuration $\hat{\mathcal{B}}$ has been determined as described in 
Sec.~\ref{inverseProblemSection}, any subsequent loading application which
we want to term {\it "forward problem"} can be dealt with computing the total deformation gradient according
to $\ba F = \bar{\ba F}\hat{\ba F}$ where $\bar{\ba F}$ is a function of 
the displacement increment $\bar{\ba u}$ due to some additional loading
and $\hat{\ba F}$ remains constant.
In practise, when employed to a numerical method, $\hat{\ba F}$ needs to be stored 
at all points of interest, e.g. numerical integration points.

Now, for forward problem the original variational formulation 
(Eq.~\ref{standardVariationalPrinciple}) is reformulated with $dV = \hat{J}^{-1}d\hat{v}$ 
as follows:
\beq
\int_{\hat{\mathcal{B}}}\,\hat{J}^{-1}\ba F\ba S\hat{\ba F}^T:\delta\bar{\ba F}\,d\hat{v} 
-\mathcal{W}_{ext} = 0
\eeq
where the variation of $\ba F$ expressed as
\beq
\delta\ba F = \delta\bar{\ba F}\hat{\ba F}\,.
\eeq
For the latter it is important to realize that, as $\hat{\ba F}$ is constant during the 
subsequent "forward" computation, the variation is only performed with respect 
$\bar{\ba u} = \ba x - \hat{\ba x}$, as $\bar{\ba F} = \bar{\ba F}(\bar{\ba u})$
is linking $\hat{\mathcal{B}}$ and  $\mathcal{B}_t$. 

With regards to deformation dependent loading we consider
\beq
\mathcal{W}_{ext} = 
\int_{\partial\hat{\mathcal{B}}_N}\,p\,\bar{J}\bar{\ba F}^{-T}\hat{\ba n}\cdot\delta\bar{\ba u}\,d\hat{a}
\label{forwardExternalPressure}
\eeq
using {\it Nanson's} formula $\bar{\ba n}\,d\bar{a} = \bar{J}\bar{\ba F}^{-T}\hat{\ba n}\,d\hat{a}$
which establishes the relation between unknown deformed and corresponding known deformed surface 
elements $d\bar{\ba a}$ and $d\hat{\ba a}$, respectively.
For deformation independent traction loading (Eq.~\eqref{inverseExternalTraction}) still applies. Consequently, for the forward problem, as for the standard total Lagrangian approach, the pressure contribution (Eq.~\eqref{forwardExternalPressure}) is to be linearized with respect to $\bar{\ba u}$ whereas the linearization of the surface traction contribution
does vanish, as $\hat{\ba F}$ is now constant.

\section{Inverse parameter optimisation method}
\label{sec:Inverse parameter optimisation method}

As previously mentioned, the determination of elastic myocardial material parameters requires knowledge of ventricular filling pressure and corresponding deformation. The latter can include deformation metrics such as change of cavity volume, short-axis and long-axis cavity expansion, twisting about the long-axis or globally and regionally averaged myocardial strains, e.g. \cite{genet2014distribution,palit2018vivo,sack2018construction}. 

Here, we aim at determining the elastic myocardial material parameters of biventricular (BV) models without prior knowledge of ventricular filling pressure data by exploiting the characteristic exponential nature of the "Klotz curve" fitted to three patient-specific left ventricular volume data, namely ESV and EDV from Cardiovascular Magnetic Resonance (CMR) imaging as well as the undeformed cavity volume, $V_0$, as obtained by inverse computation using the non-standard total Lagrangian approach presented in Sec.~\ref{nonstandardTheorySection} implemented in a finite element method (FEM)-based code. 

This is achieved by iterative parameter optimisation employing the Levenberg-Marquardt algorithm (LVM) \cite{Guyon2000least}. Further details regarding the specific LVM implementation can be found in \cite{rama2019real}. This iterative scheme is initialized by assuming a physiological EDP and elastic myocardial material parameters for the LV which serves to construct an initial volume-normalized "Klotz curve" \cite{klotz2006single}. Before elaborating on the details of the proposed iterative scheme, the important relations needed to set-up the "Klotz curve" are briefly summarized. Given a measured pair of LV cavity volume, $V_m$, and corresponding cavity pressure, $P_m$, the unloaded LV cavity volume can be estimated as
\begin{equation}
V_0 = V_m\,\left(0.6-0.006\,P_m\right)
\end{equation}
and at 30\,\text{mmHg} as
\begin{equation}
V_{30} = V_0 + \frac{V_{m,n}-V_0}{(P_m/A_n)^{(1/B_n)}}
\end{equation}
where the normalized measured volume is given by
\begin{equation}
V_{m,n} = \frac{V_m-V_0}{V_{30}-V_0}
\end{equation}
and for humans the coefficients $A_n = 27.8$ and $B_n = 2.76$. Then the EDPVR is formulated as
\begin{equation}
P_{LV} = \alpha\,V_{LV}^\beta    
\label{eq:Klotz curve}
\end{equation}
with 
\begin{equation}
\alpha = \displaystyle\frac{30}{\left(V_{30}\right)^{\text{Log}(P_m/30)/\text{Log}(V_m/V_{30})}}
\end{equation}
and
\begin{equation}
\beta = \displaystyle\frac{\text{Log}(P_m/30)}{\text{Log}(V_m/V_{30})}\,.
\end{equation}
Now, in the absence of LV pressure measurements, EDP is to be iteratively estimated taking $P_m = \text{EDP}$ and $V_m = \text{EDV}$. For the initial setup of the "Klotz curve", a physiologically reasonable EDP value needs to be chosen. ESP and $V_0$ can then be readily obtained via Eq.~\eqref{eq:Klotz curve} knowing ESV from CMR and $P_0 = 0$. The LVM cost function considers the fitting error of $V_0$ and EDV between "Klotz curve" and the FEM solution. As cavity pressure data are generally not available for the RV, it is therefore assumed that the "Klotz curve" (Eq.~\eqref{eq:Klotz curve}) can provide an indication of the EDPVR of the RV as well. Due to this limitation, the LVM cost function only includes the error regarding EDV but not $V_0$ for the RV. For each subsequent LVM iteration, updated elastic parameter and EDP values are obtained for LV and RV from the optimisation algorithm allowing for the update of the "Klotz curve" and subsequently ESP and $V_0$ values as well. A noteworthy aspect of this approach is that the fitting target, which is the "Klotz curve", is continuously changing which is not the case in the standard application of LVM. The iterative parameter optimisation is terminated once the cost function error falls below a given threshold and the root-square error of the entire EDPVR of "Klotz curve" and FE simulation is smaller than 2.00 mmHg.


\begin{figure}[h]
	\centering
	\includegraphics[height=125mm]{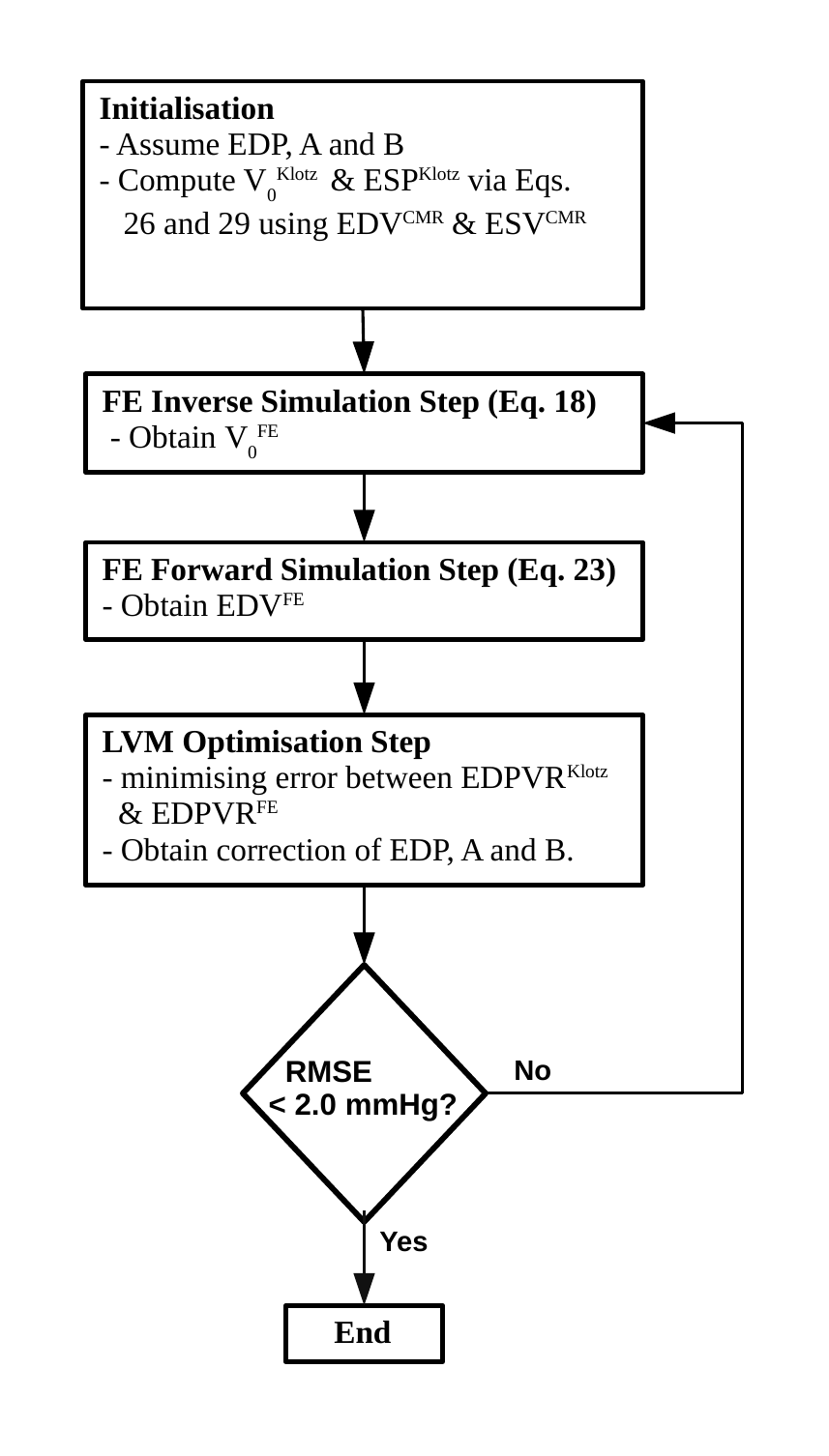}
		\vspace*{-0.5cm}
	\caption{Inverse procedure combining "Klotz Curve", FE simulation and Levenberg-Marquardt (LVM) optimization schemes.\label{figa1}}
\end{figure}


\section{Numerical examples\label{examplesSection}}

The following examples presented in this section aim to demonstrate the potential of the proposed inverse parameter optimisation method outlined in Sec.~\ref{sec:Inverse parameter optimisation method}. In this study, only the diastolic filling of the biventricular (BV) cavities will be examined.

\begin{table}[h]
    \centering
     \caption{Ventricular cavity volumes measured by CMR.}
    \begin{tabular}{|c|c|c|c|c|c|c|}
    \hline
    & \multicolumn{3}{|c|}{Left ventricle}  &  \multicolumn{3}{|c|}{Right ventricle} \\ \hline
     & ESV & EDV & SV & ESV & EDV & SV \\ \hline
    Case 1 & 90.00 & 168.00 & 78.00 & 61.00 & 96.00 & 78.00 \\
    Case 2 & 86.00 & 170.00 & 84.00 & 67.00 & 118.00 & 51.00 \\
    Case 3 & 82.00 & 157.00 & 75.00 &78.00 & 160.00 & 82.00 \\ \hline
    \end{tabular} \\
    \label{tab1}
\end{table}

We constructed patient-specific heart models of three healthy patients that were scanned using Cardiovascular Magnetic Resonance (CMR) imaging at the Groote Schuur Hospital, University of Cape Town, South Africa.  University of Cape Town, Faculty of Health Sciences Human Research Ethics Committee approval (REF: 686/2018) and patients’ consent were obtained to conduct research on unidentified human data. The left and right ventricular cavity volumes at end-systole and end-diastole, respectively, are presented in Tab.~\ref{tab1}. The anatomical models were meshed using 3326, 2765 and 3246 linear tetrahedral elements, for the cases 1, 2, and 3, respectively. ESP cavity filling pressure boundary conditions were applied to the endocardial surfaces of LV and RV according to  Eq.~\eqref{intermediateExternalPressure} computing first the unloaded configuration as outlined in Sec.~\ref{inverseProblemSection} and EDP cavity filling pressure boundary conditions according to Eq.~\eqref{forwardExternalPressure} computing subsequently the end diastolic configuration as outlined in Sec.~\ref{forwardProblemSection}.
In terms of the Dirichlet boundary conditions, the surface of the heart's base is fixed to restrain vertical direction movement. To allow for torsional behaviour and wall thickening and thinning, a Dirichlet boundary condition is weakly imposed through application of an elastic line force of $0.1kN$ acting in tangential direction around the epicardial base. The elastic line forces effectively prevent rigid body motion in short-axis directions as well.

The material behaviour of the BV heart muscle tissue is assumed to be nonlinear, anisotropic and nearly incompressible described by the following orthotropic strain energy function \citet{usyk2000effect}

\begin{small}
\beq
\psi &=& \frac{A}{2} \left( \exp^{BQ} - 1 \right)  + A_{\text{comp}} 
	\Big( \text{det} J \; \text{ln}\left( \text{det} J \right) -  \text{det} J + 1 \Big),
%
\eeq
\end{small}
where parameters $A$ and $B$ are stiffness factors and $J = \text{det} \ba F$ is the Jacobian quantifying volumetric deformation of cardiac tissue is linked to the constant $A_{comp}$ controlling the compressibility of the myocardium. Additionally, the exponent $Q$ is defined in terms of the Green strain tensor and the material directions defining structural tensors $\ba M_f$, $\ba M_s$ and $ \ba M_n$ as follows \cite{rama2016real}:
\beq
Q &:=& a_1(\text{tr}(\ba M_f \ba E))^2 + a_2(\text{tr}(\ba M_s \ba E))^2 + a_3(\text{tr}(\ba M_n \ba E))^2 \nonumber\\
& & + a_4\,\text{tr}(\ba M_f \ba E^2) + a_5\,\text{tr}(\ba M_s \ba E^2) + a_6\,\text{tr}(\ba M_n \ba E^2),
\eeq
where $a_i,\,i = 1, ..., 6$ are the anisotropy coefficients associated with the three preferred material directions, namely fiber axis, $\ba V_f$, sheet axis, $\ba V_s$, and sheet normal axis, $\ba V_n$. These vectors construct an orthonormal basis and allow for the formulation of the so-called structural tensors
\beq
\ba M_f = \ba V_f \otimes \ba V_f, \ba \ \ M_s = \ba V_s \otimes \ba V_s, \ \ \ba M_n =  \ba V_n \otimes \ba V_n.
\eeq
The values of the anisotropy material constants used are listed in Table ~\ref{tab2}.
\begin{table}[h]
    \centering
    \caption{Anisotropy coefficients used.}
    \begin{tabular}{|c|c|c|c|c|c|c|}
    \hline
        $A_{comp}$  (kPa) & $a_1$ & $a_2$ & $a_3$ & $a_4$ & $a_5$ & $a_6$ \\ \hline
       100 & 6 & 7 & 3 & 12 & 3 & 3 \\ \hline
    \end{tabular}
        \label{tab2}
\end{table}

\begin{figure}[h]
\centering
	\includegraphics[width=8.5 cm]{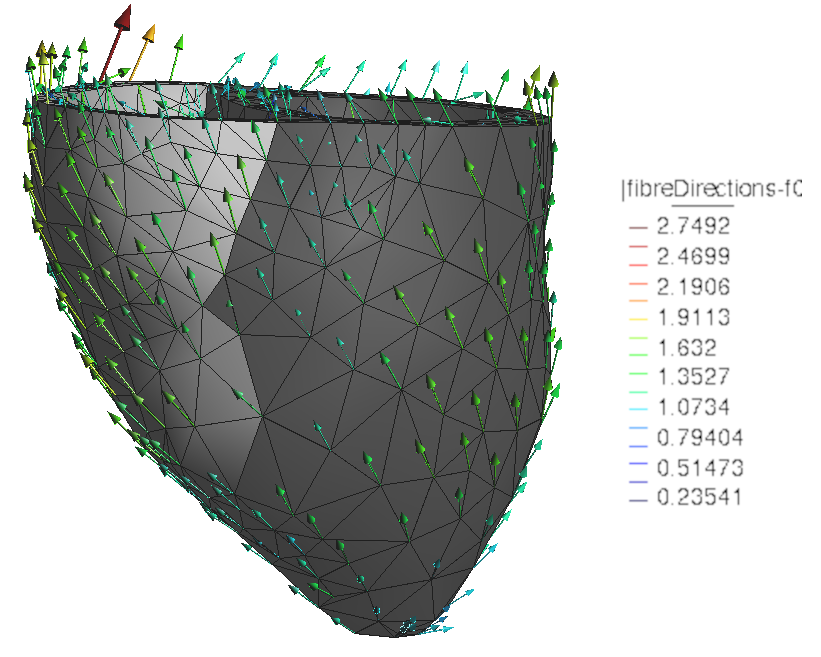}
	\caption{Myocardial fibre orientation of a bi-ventricular model.\label{fibre}}
\end{figure}

The varying fibre directions throughout the LV and RV walls are computed adopting an algorithm developed by Wong and Kuhl \cite{wenk2011novel} but using moving least square (MLS) based approximations as proposed by \citet{skatulla2016path} instead of the solving a Poisson problem. The resulting fibre direction distribution is shown for Case 2 in Fig.~\ref{fibre}. The prescribed fibre direction angles at the endocardium and epicardium of both LV and RV are $+72^o$ and $-57^o$, respectively \cite{rama2019real}.

\subsection{Iterative parameter estimation}

\begin{figure}[h]
	\includegraphics[width=9 cm]{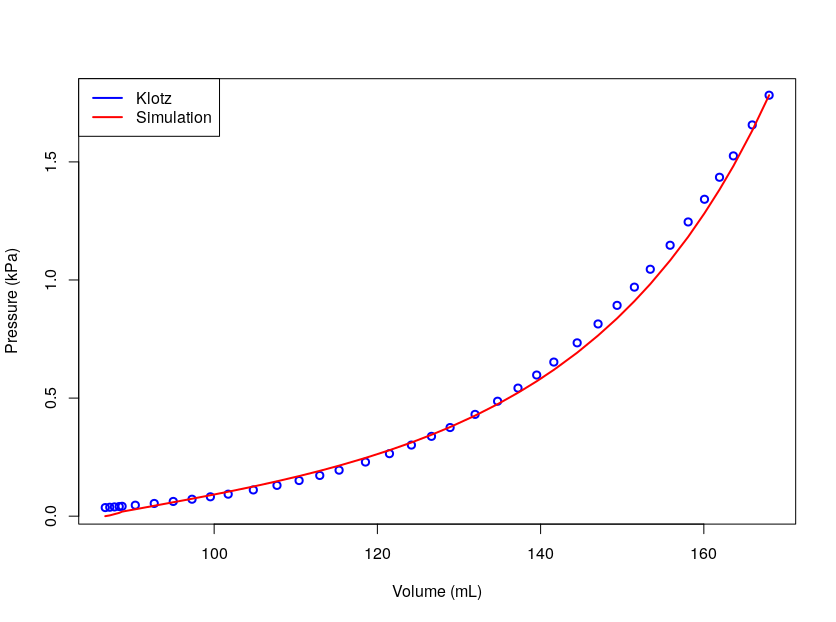}
	\caption{EDPVR curve predicted by Usyk model (red) and method of Klotz et al. (blue) as a physiological benchmark for case 1.\label{fig1}}
\end{figure}

\begin{figure}[h]
	\includegraphics[width=9 cm]{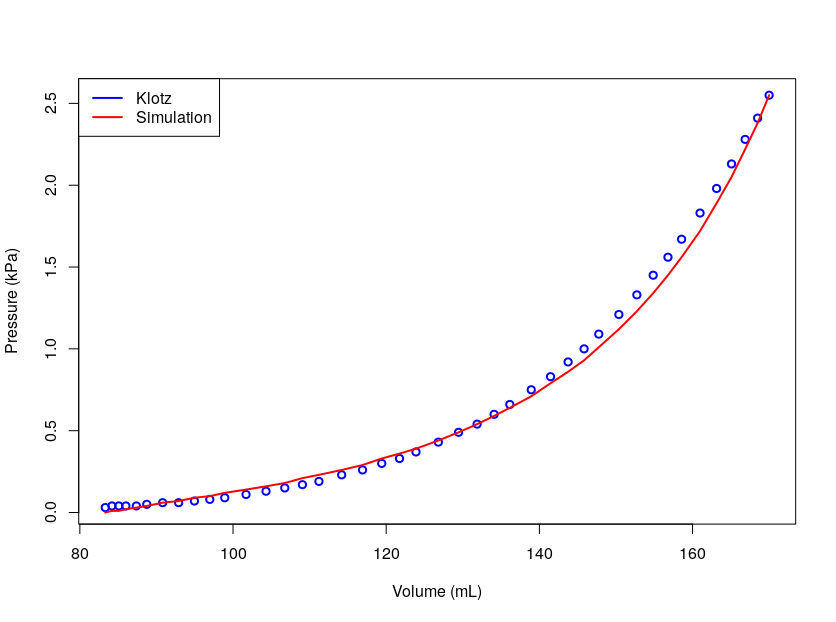}
	\caption{EDPVR curve predicted by Usyk model (red) and method of Klotz et al. (blue) as a physiological benchmark for case 2.\label{fig2}}
\end{figure}

\begin{figure}[h]
	\includegraphics[width=9 cm]{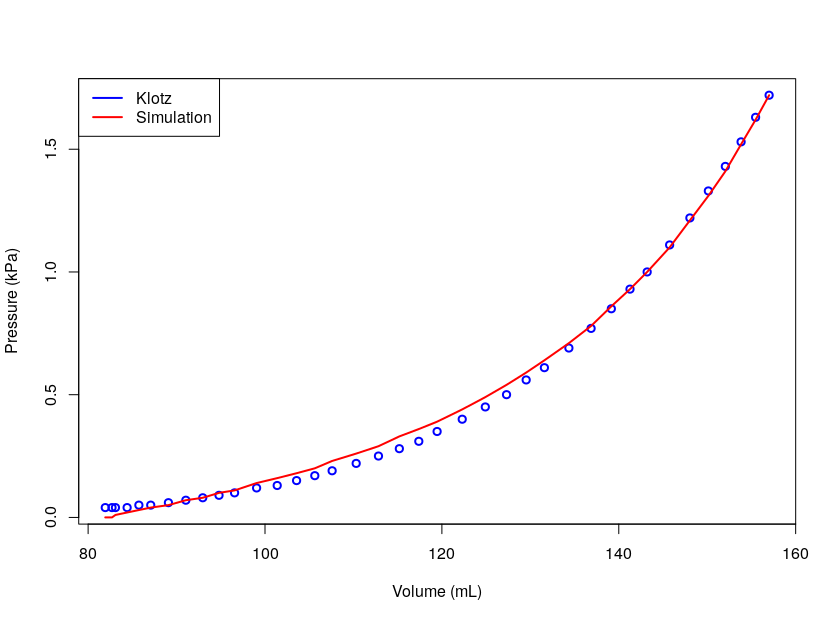}
	\caption{EDPVR curve predicted by Usyk model (red) and method of Klotz et al. (blue) as a physiological benchmark for case 3.\label{fig3}}
\end{figure}

The iterative parameter estimation approach is used to obtain patient-specific values for the characteristic cavity pressures $EDP$ and $ESP$, respectively, as well as the stiffness parameters $A$ and $B$ for LV and RV. The corresponding physiologically motivated initial values are listed in Tab.~\ref{tab:initial values}. The LVM scaling parameters are set as $\lambda^{\text{LVM}} = 1$ and $\nu^{\text{LVM}} = 2$, respectively and the stopping criterion of the cost function residual is chosen as $\gamma^\text{LVM}_\epsilon = 1 X 10^{-2} $. 
\begin{table}[h]
    \centering
     \caption{Initial material parameter and cavities pressure values in kPa \cite{zhou2020clinical,mielniczuk2007left,wang2013structure}}.
    \begin{tabular}{|c|c|c|c|c|c|c|c|c|}
    \hline
    & \multicolumn{3}{c}{Left ventricle}  &  \multicolumn{3}{|c|}{Right ventricle} \\ \hline
     & A (kPa) & B  &$EDP$ & A (kPa) & B &  $EDP$ \\ \hline
    Case 1 & 0.10&	1.00&		2.00&	0.10&	1.00&		0.50\\
    Case 2 & 0.10&	1.00&		3.00&	0.10&	1.00&		1.50 \\
    Case 3 & 0.10&	1.00&		2.00&	0.10&	1.00&		2.50 \\ \hline
    \end{tabular} \\
    \label{tab:initial values}
\end{table}
The final obtained stiffness parameter and characteristic pressure values are listed in Tab.~\ref{tab3}. 
\begin{table}[h]
    \centering
     \caption{Final estimated material parameters and cavities pressure values in kPa.}
    \begin{tabular}{|c|c|c|c|c|c|c|c|c|}
    \hline
    & \multicolumn{4}{c}{Left ventricle}  &  \multicolumn{4}{|c|}{Right ventricle} \\ \hline
     & A (kPa) & B & $ESP$ &$EDP$ & A (kPa) & B & $ESP$ & $EDP$ \\ \hline
    Case 1 & 0.10&	1.25&	0.04&	1.78&	0.11&	1.05&	0.08&	0.85 \\
    Case 2 & 0.09&	1.20&	0.04&	2.55&	0.12&	1.21&	0.04&	1.06 \\
    Case 3 & 0.09&	1.07&	0.04&	1.72&	0.10&	0.88&	0.02&	1.25 \\ \hline
    \end{tabular} \\
    \label{tab3}
\end{table}
The converged iterative parameter optimisation scheme gave for the unloaded LV cavity volumes $V_0 = 86.66,\, 83.32\,\text{and}\,81.94\,\text{mL}$ and for the unloaded RV cavity volumes $V_0 = 53.92,\, 65.18\,\text{and}\,86.99\,\text{mL}$ for cases 1, 2 and 3, respectively.
Figs.~\ref{fig1} - ~\ref{fig3} show the fitted EDPVR curves illustrating the FEM solutions by red lines and the targeted "Klotz curve" by blue circles indicating good agreement. The root square mean error (RMSE) are 0.225, 0.225 and 0.375 mmHg for cases 1, 2 and 3 respectively at the LV and 0.150, 0.150 and 0.075 mmHg also for cases 1, 2 and 3 respectively. 

The average percentage difference between the simulated and targeted CMR LVEDVs was approximately 0.0007$\%$, and 0.0020$\%$ for the RVEDVs. Similarly, the average percentage difference between the simulated and targeted Klotz $V_0$ for the LV was approximately 0.00005$\%$ and 5.3314$\%$ for the RV. The observed higher percentage error in the RV initial volume confirm that we did not target RV initial volume in our calibration. For the segmented and CMR ESV, there is 0.92$\%$ average error difference at the LV and 3.00$\%$ at the RV, detailed result is on Table. ~\ref{tab4}. 

\begin{table}[h]
    \centering
    \caption{Percentage error of LV and RV volumetric targets and results from the calibration.}
    \begin{tabular}{|c|c|c|c|c|c|c|}
    \hline
    & \multicolumn{3}{c}{Left ventricle}  &  \multicolumn{3}{|c|}{Right ventricle} \\ \hline
         & $V_0$  & LVEDV  & LVESV & $V_0$  & RVEDV  & RVESV   \\ \hline
       Case 1 & 0.00005 &	0.00172 &	1.42 & 1.73 &	0.0003 & 3.92 \\
       Case 2 & 0.00006 &	0.00029 &	0.01 & 0.01 &	0.0020 &	1.86 \\
       Case 3 & 0.00002 &	0.00005 &	1.33 & 14.25 &	0.0036 &	3.22     \\\hline       
    \end{tabular}
      \label{tab4}
\end{table}

\subsection{Stress and strain distributions}

\begin{figure}[h]
\centering
 \includegraphics[width=12.5 cm]{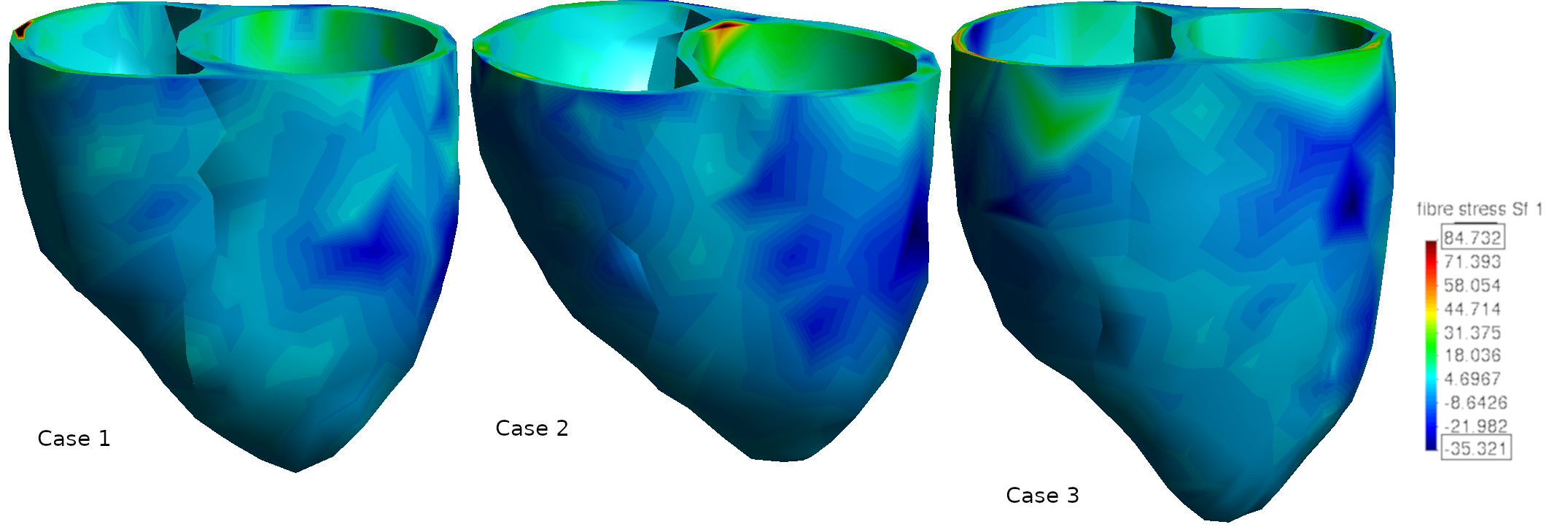}
	\caption{Fibre stress at the end-diastole.\label{fig5}}
\end{figure}

\begin{figure}[h]
\centering
 \includegraphics[width=12.5 cm]{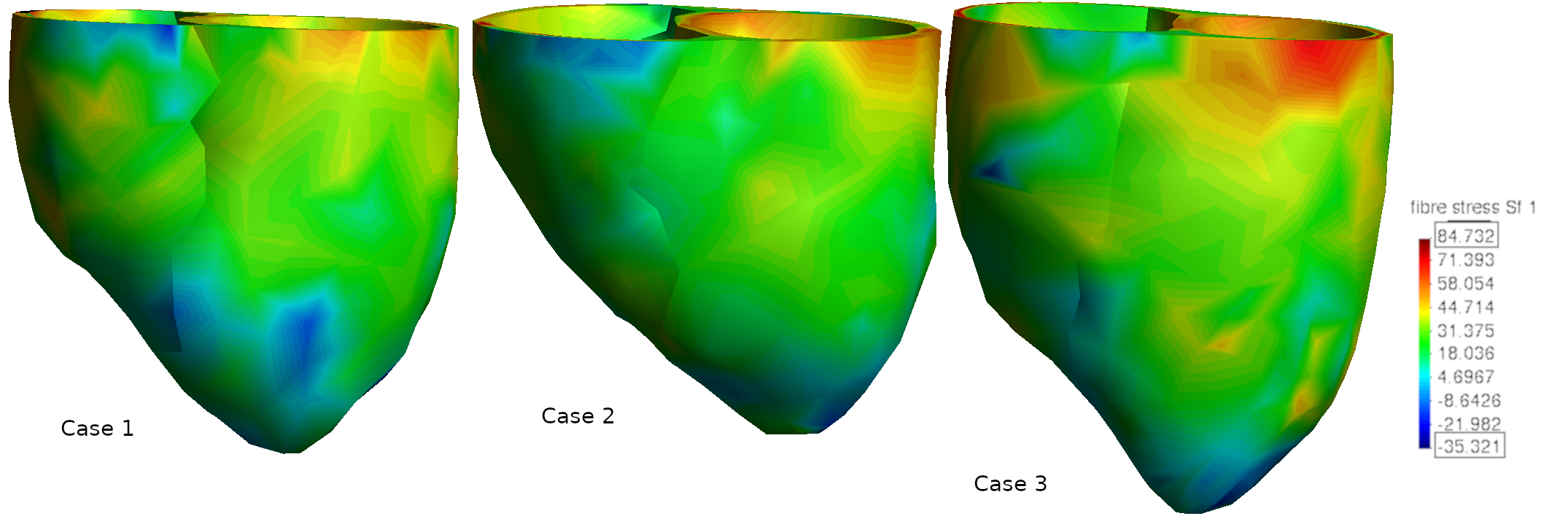}
	\caption{Fibre strain at the end-diastole.\label{fig6}}
\end{figure}

The contour plots in Figs.~\ref{fig5} and ~\ref{fig6} reveal high myofibre stress and strain concentrations at the base which is to be expect due to the geometrical support. 
The global residual fibre stress, separately averaged over LV and RV myocardium, at end systole is found as 0.013, 0.018 and 0.006 kPa for LV at the longitudinal, circumferential and radial respectively and 0.031, 0.038, and 0.012 kPa for RV longitudinal, circumferential and radial respectively.
A detailed analysis of regional strain in the LV with respect to the global longitudinal, circumferential and radial directions is presented. Endocardial strain comparison demonstrated moderate agreement of global strains, separately averaged over the LV and RV myocardium, respectively. Global longitudinal strain values (GLS) for cases 1 to 3 are 5.75, 6.90 and 7.14$\%$ for the FE model simulation, while the CMR in vivo recorded values for cases 1 to 3 are given as  12.30, 17.80 and 21.60$\%$ respectively. The global circumferential strain values (GCS) for the FE simulations are 22.2, 25.82 and 25.82$\%$, and 20.20, 20.40 and 25.70$\%$ for the in vivo. Similarly, the global radial strain values (GRS) for the FE simulations are -15.84, -17.37, and -16.39$\%$, and -13.90, -14.00, and -16.20$\%$ in vivo for the cases 1, 2 and 3 respectively.
The GCS and GRS are well comparable in both FE model simulation and the CMR in vivo, but this was not the case with longitudinal strains, whereby a regional analysis of the longitudinal strains revealed poor agreement between simulated and measured in vivo strains. Although discrepancies have been noted between the FEM and in vivo longitudinal strain \cite{sack2018construction} due to  the absence of data on the true distribution of the patient-specific myofiber orientation angles, which affects the circumferential-longitudinal compliance ratio. Also the CMR long-axis scan providing longitudinal strain and the short-axis scan providing circumferential and radial strains are obtained from different cardiac cycles whereas the FEM calculate these strains using the same cardiac cycle.


\section{Conclusion}
In this paper, a direct inverse method to determine the unloaded configurations of biventricular patient-specific heart models is combined with the bounded Levenberg-Marquart parameter optimisation method to estimated besides elastic material parameters also the unknown end-systolic and end-diastolic ventricular pressures in silico. The unloaded configuration is computed by a modified total Lagrangian approach as implemented in standard FEM which implicitly provides for the incorporation of pres-stressing and pre-straining of the myocardium as linked to the end-systolic pressure. The proposed variational principle is derived from the standard Lagrangian formulation, in contrast to the similar inverse unloaded configuration determination methods by \citet{govindjee1996computational} and \citet{gee2010computational}. The unknown characteristic ventricular pressures and elastic material parameters can then be iteratively computed by taking advantage of the highly nonlinear exponential nature of the EDPVR and fitting the patient-specific LV EDPVR to the analytical volume-normalized "Klotz EDPVR curve". Making use of CMR-derived anatomical heart models, the iterative framework therefore provides the means to non-invasively estimate LV pressure, pre-strain and elastic material parameters exploiting their biomechanical interdependence.  

Our results indicate good agreement between the EDPVR curves obtained from FE simulation and given by the "Klotz Curve". Additionally, there is a close agreement in the FE global circumferential and radial strains when compared with the CMR global strain data. The global longitudinal strain shows poor agreement between FE and CMR measured strains. This discrepancy has been attributed to the absence of data on the true distribution of the patient-specific myofiber orientation distribution, which affects the circumferential-longitudinal compliance ratio \cite{sack2018construction}. Also the CMR long-axis scan providing longitudinal strain and short-axis scan providing circumferential and radial strains are obtained from different cardiac cycles whereas FEM calculates these strains within the same cardiac cycle. Another limitation of this study is the absence of in vivo pressure measurements to validate our results. In future work, we shall obtain pressure data in vivo to perform validation of our proposed method, also patient-specific fiber orientation distribution data from DTMRI will help to mitigate the error in FE and CMR global longitudinal strains.

\section*{References}

\begingroup
\bibliographystyle{plainnat}
\renewcommand{\section}[2]{}
\bibliography{references}
\endgroup

\end{document}